\documentstyle[preprint,aps,epsfig]{revtex}
\begin{document}
\preprint{\font\fortssbx=cmssbx10 scaled \magstep2
\hfill$\vcenter{\hbox{\bf CERN-TH/95-80}
                \hbox{\bf hep-ph/9504246}}$}
\title{Future $\nu_\tau$ Oscillation Experiments and Present Data}
\author{J.\ J.\ Gomez-Cadenas
\footnote{E-mail: jjgomez@phast.umass.edu or jjgomez@huhepl.harvard.edu.
On leave from PPE Division, CERN,
CH-1211 Geneva 23, Switzerland,
and Departamento de F\'{\i}sica At\'omica y Nuclear,
Universidad de Valencia, Spain}}
\address{Department of Physics and Astronomy, University of Massachusetts \\
 Amherst, MA 01003, USA.}
\author{M.\ C.\ Gonzalez-Garcia
\footnote{E-mail: concha@vxcern.cern.ch (internet) or
VXCERN::CONCHA (decnet).}
}
\address{Theory Division,   CERN,
CH-1211 Geneva 23, Switzerland.}
\maketitle
\thispagestyle{empty}
\begin{abstract}
\baselineskip 0.42 cm
Our goal in this paper is to examine the discovery potential of laboratory
experiments searching for the oscillation $\nu_\mu(\nu_e) \rightarrow
\nu_\tau$, in the light of recent data on solar and atmospheric neutrino
experiments, which we analyse together with the most restrictive results from
laboratory experiments on neutrino oscillations. In order to explain
simultaneously $all$ present results we use a four-neutrino framework, with an
additional sterile neutrino. Our predictions are rather pessimistic for the
upcoming  experiments NOMAD and CHORUS, which, we find, are able  to explore
only a small area of the oscillation parameter space. On the other hand, the
discovery potential of future experiments is much larger. We consider three
examples. E803, which is approved to operate in the future Fermilab main
injector beam line, MINOS, a proposed long-baseline  experiment also using the
Fermilab beam, and NAUSICAA, an improved detector which improves by an
order of magnitude the performance of CHORUS/NOMAD and can be
operated either at CERN or
at Fermilab beams. We find that those experiments can cover a very substantial
fraction of the oscillation parameter space, having thus a very good chance of
discovering $both$ $\nu_\mu \rightarrow \nu_\tau$ and $\nu_e \rightarrow
\nu_\tau$ oscillation modes.
\end{abstract}
\noindent
{\bf CERN-TH/95-80} \\
\noindent
\vskip -1. cm
{\bf April 1995}
\newpage
\section{Introduction}
Present data from solar and atmospheric neutrino experiments favour the
hypothesis of neutrino oscillations. Nevertheless, this interpretation requires
confirmation from further experiments, in particular from laboratory
experiments. All  solar neutrino experiments \cite{solar} find less $\nu_e$
than predicted theoretically. However, the uncertainties in the calculation of
the solar neutrino flux may still be rather large \cite{solflux} and new
experiments  \cite{newsolar,newat} are being planned to further investigate the
possible origin of the solar neutrino deficit. As for  atmospheric neutrino
experiments, two of them \cite{kamisub,kamimul,IMB} measure a ratio
$\nu_\mu/\nu_e$ smaller than expected from theoretical calculations. Here, in
addition to the uncertainties on the estimation of
the atmospheric neutrino flux  and on the neutrino-nucleon cross section
\cite{paolo}, one has to add the uncertainties due to the modest data sample.
New experiments are also being planned in this area \cite{newat}.

A complementary approach is to look for neutrino oscillations in laboratory
experiments, where the experimental conditions, in particular the shape,
energy, and flux of the neutrino beam are under control. A number of neutrino
experiments have recently started taking data. These are the LSND experiment
at Los Alamos \cite{LSNDexp}, which looks for $\nu_\mu \rightarrow \nu_e$
oscillations, and the CHORUS \cite{chorus} and NOMAD \cite{nomad}  experiments
at CERN, which are primarily sensitive to $\nu_\mu (\nu_e) \rightarrow
\nu_\tau$ oscillations. Recent data from LSND may be consistent with the
existence of neutrino oscillations \cite{LSND}, although no formal claim has
been made so far by the collaboration. First results from CHORUS and NOMAD
should be available in 1996. In addition, a number of new $\nu_\mu(\nu_e)
\rightarrow \nu_\tau$ oscillation experiments are being discussed.  At
Fermilab, a new, very intense $\nu_\mu(\nu_e)$ beam is planned to be available
around the year 2001. Two complementary experiments are being discussed to
benefit from this beam. These are E803 \cite{E803}, a short-baseline
experiment similar in design to CHORUS, and MINOS \cite{minos}, a long-baseline
experiment, which proposes to detect the neutrinos produced at Fermilab with a
$10$ Kton detector located in the Soudan mine, around 730 km away from the
neutrino source. Furthermore, new experimental techniques are being studied to
upgrade NOMAD and CHORUS \cite{nomad01,capilar}. Ultimately, these new
techniques could result in the design of new detectors able to improve the
performance of NOMAD and CHORUS/E803 by one order of magnitude
\cite{carlo,klaus,nomad01}. Such  detector(s) could be operated either at CERN
or at Fermilab beams.

In this paper we examine the prospects of success of all these
experiments in the light of present data. CHORUS and NOMAD were conceived at a
time when the dominant scenario for neutrino masses was consistent with very
light $\nu_e$, $\nu_\mu$ and a $\nu_\tau$ of about $10$ eV constituting the hot
component of the dark matter. This scenario arose as the most natural solution
to explain the solar neutrino problem,  simultaneously providing a
candidate for the hot dark matter. However, present data no
longer favour this simple scenario. Indeed, explaining the results from solar
and atmospheric neutrino experiments in the usual three-neutrino framework is
very difficult and requires a large degree of fine-tuning. One has to choose
between throwing away part of the data and considering a larger scheme. The
``minimal" scheme to explain $all$ data without fine-tuning seems to be
a four-neutrino framework
$(\nu_e,\nu_\mu,\nu_\tau,\nu_s)$ where $\nu_s$ is a sterile neutrino.
Using this framework we perform
a consistent analysis of the data from solar and atmospheric neutrino
experiments as well as  results on neutrino-oscillation laboratory experiments.
This analysis enables us to re-assess the discovery potential of CHORUS and
NOMAD as well as to study the prospects for the new experiments discussed
above. We find that CHORUS and NOMAD have a rather marginal chance of
discovering $\nu_\mu(\nu_e) \rightarrow \nu_\tau$ oscillations. However,
the future $\nu_\tau$ oscillation experiments have much better prospects
since they cover a very large fraction of the oscillation parameter space.

The outline of the paper is as follows. In Sec.\ \ref{for} we review the
formalism for neutrino oscillation in a general multi-family framework.
Section \ref{exp} is devoted to a summary of the present experimental status
for
solar and atmospheric neutrino experiments as well as accelerator and reactor
neutrino oscillation experiments.  The basic ingredients of the four-neutrino
framework are presented in Sec.\ \ref{4famil} and the results of  our
analysis of the  available experimental data, discussed in Sec.\ \ref{exp},
are displayed in Sec.\ \ref{global}. Section \ref{experiments} describes
succinctly the upcoming and future $\nu_\tau$ oscillation experiments while
Sec.\ \ref{discovery} is dedicated to the study of the prospects for the
discovery of $\nu_\tau$ oscillations.  Finally we present our conclusions in
Sec.\ \ref{conclu}.


\section{Formalism}
\label{for}
If neutrinos have a mass, the weak eigenstates $\nu_\alpha$ produced in a
weak interaction (i.e., an inverse beta reaction or a weak decay)
will be, in general, a linear combination of the mass eigenstates $\nu_i$
\begin{equation}
\nu_\alpha =\sum_{i=1}^{n} U_{\alpha i} \nu_i \;
\end{equation}
where $n$ is the number of neutrino species and $U$ is a unitary mixing matrix.

After travelling a distance $L$, the neutrino can be
detected in the charged-current (CC) interaction $\nu \; N' \rightarrow
l_\beta \; N $
with a probability
\begin{equation}
P_{\alpha\beta}=\delta_{\alpha\beta}-4\sum_{i=1}^n\sum_{j=i+1}^n
\mbox{Re}[ U_{\alpha i}U^\star_{\beta i} U^\star_{\alpha j} U_{\beta j}]
\sin^2\left(\frac{\Delta_{ij}}{2}\right)
\end{equation}
The probability, therefore, oscillates with oscillation lengths $\Delta_{ij}$
given by
\begin{equation}
\frac{\Delta_{ij}}{2}=1.27 \frac{|m_i^2-m_j^2|}{\mbox{eV}^2}
\frac{L/E}{\mbox{m/MeV}}=
1.27 \frac{\Delta m^2_{ij}}{\mbox{eV}^2} \frac{L/E}{\mbox{m/MeV}}
\end{equation}
where $E$ is the neutrino energy.
Each experimental set up has a different characteristic value of the ratio
$L/E$ and is thus most sensitive
to oscillation lengths such  that
$\Delta m^2_{ij} \approx 1/(L/E)$. The typical values of $L/E$  are
summarized in Table \ref{tab:lovere}.

In general neutrino beams are not monochromatic. Thus, rather than measuring
$P_{\alpha \beta}$, the experiments are sensitive to the average probability
\begin{equation}
\begin{array}{ll}
\langle P_{\alpha \beta}\rangle &= \frac{\displaystyle \int dE_\nu
\frac{d\Phi}{d E_\nu}
\sigma_{CC}(E_\nu) P_{\alpha \beta} (E_\nu) \epsilon(E_\nu)}
{\displaystyle \int dE_\nu  \frac{d\Phi}{d E_\nu}
\sigma_{CC}(E_\nu)  \epsilon(E_\nu)} \\
& = {\displaystyle \delta_{\alpha,\beta}-4\sum_{i=1}^n\sum_{j=i+1}^n
\mbox{Re}[ U_{\alpha i}U^\star_{\beta i} U^\star_{\alpha j} U_{\beta j}]
\langle\sin^2\left(\frac{\Delta_{ij}}{2}\right)\rangle },
\end{array}
\end{equation}
where $\Phi$ is the neutrino energy spectrum, $\sigma_{CC}$ is the cross
section for the process in which the neutrino is detected (in general a CC
interaction), and  $\epsilon(E_\nu)$ is the detection efficiency for the
experiment.
For oscillation lengths such that $\Delta m^2_{ij} \gg 1/(L/E)$ the oscillating
phase will have been over many cycles before the detection and therefore
it will have averaged to $\langle \sin^2(\Delta_{ij}/{2})\rangle ={1}/{2}$.
On the other hand, for $\Delta m^2_{ij} \ll 1/(L/E)$, the oscillation did not
have time to give any effect and  $\langle \sin^2({\Delta_{ij}}/{2})\rangle
\approx 0$

For the case of two-neutrino oscillations the above expressions take the well
known form
\begin{equation}
P_{\alpha\beta}=\delta_{\alpha\beta}- (2\delta_{\alpha\beta}-1) \sin^2
(2\theta) \langle \sin^2\left(\frac{\Delta_{12}}{2} \right) \rangle
\end{equation}
since
\begin{equation}
U=\left(\begin{array}{cc} \cos\theta & \sin\theta\\
-\sin\theta & \cos\theta \end{array} \right)
\end{equation}

Most neutrino oscillation experiments present their result in the two-family
mixing language as regions in the
plane ($\Delta m^2,\; \sin^2(2\theta)$). Using the previous expression it is
possible to translate their results into transition probabilities.


\section{Experimental Status}
\label{exp}
\subsection{Accelerator and Reactor Neutrino Experiments}
\begin{figure}
\begin{center}
\mbox{\epsfig{file=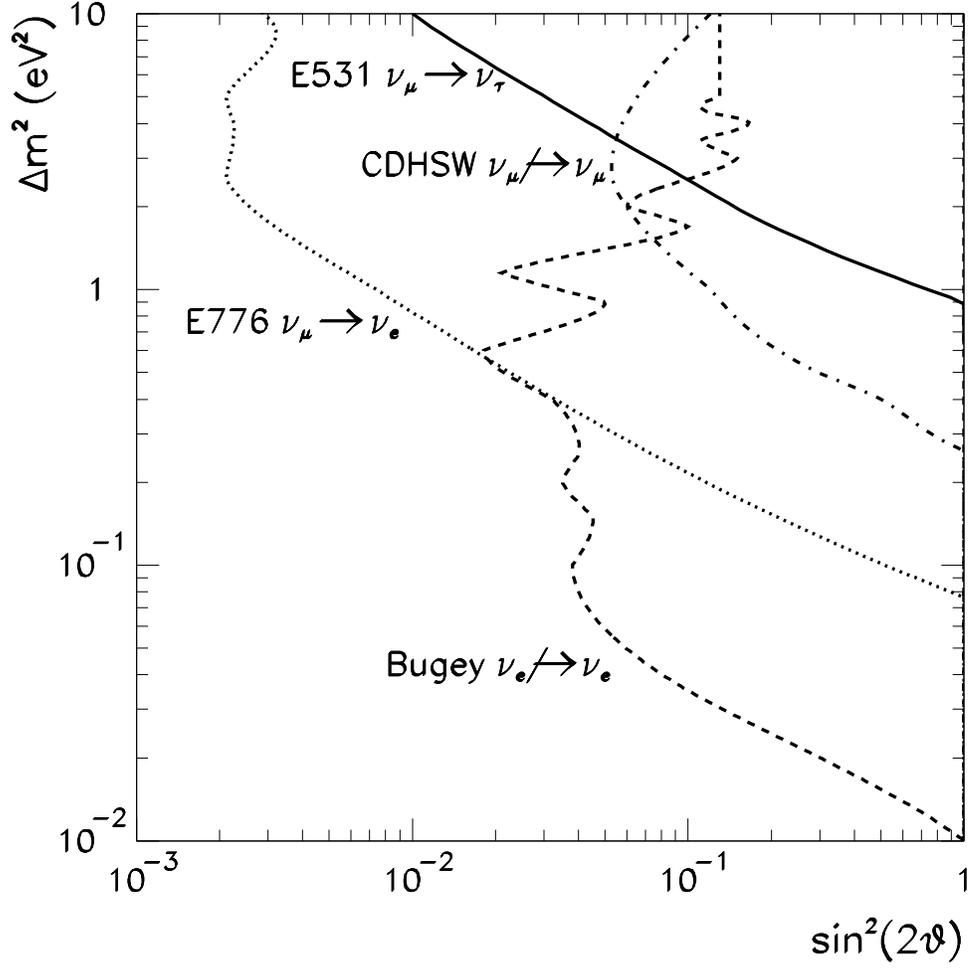,bbllx=60,bblly=200,bburx=600,bbury=600}}
\end{center}
\caption{90\% CL exclusion contours for different reactor and accelerator
experiments in the two-family mixing approximation:
E531 $\nu_\mu\rightarrow \nu_\tau$ apperarance experiment (solid line),
Bugey $\nu_e\rightarrow \!\!\!\!\!\!\! \setminus \;
\nu_e$ disappearance experiment (dashed line),
E776 $\nu_\mu\rightarrow \nu_e$ apperarance experiment (dotted line), and
CDHSW $\nu_\mu\rightarrow \!\!\!\!\!\!\!
\setminus\; \nu_\mu$ disappearance experiment
(dot-dashed line). The allowed regions lie below the curves.}
\label{2family}
\end{figure}

There are two types of laboratory experiments to search for neutrino
oscillations. In a disappearance experiment one looks for the attenuation
of a neutrino beam primarily composed of a single flavour
due to the mixing with other flavours.
At present the most restrictive experiments of this kind are
the reactor experiment at Bugey \cite{bugey}, which looks for $\nu_e$
disappearance, and the CDHSW experiment \cite{CDHSW} at CERN,  which searches
for $\nu_\mu$ disappearance.
Both experiments show no indication of neutrino oscillation. Their
results are presented as exclusion areas in the two-neutrino oscillation
approximation in Fig.\ \ref{2family}. We translate their results into
limits on the transition probabilities:
\begin{equation}
\begin{array} {l}
\langle P_{ee}\rangle  \gtrsim 0.93 \;\;\; \mbox{from Bugey for
$\Delta m^2 \gtrsim  4 $eV$^2$} \\
\langle P_{\mu\mu}\rangle  \gtrsim 0.95 \;\;\; \mbox{from CDHSW} \\
\end{array}
\end{equation}
Both results are given at $90 \%$ CL. The maximum probability from Bugey is
larger for smaller mass differences since the neutrino flux normalization can
be  better determined by the experiment. However, in the range of masses we are
interested the relevant limit is the one given above.

In an appearance experiment one searches for interactions by neutrinos of a
flavour not present in the neutrino beam. The most restrictive experiments  are
the E776 experiment at BNL \cite{E776}  for the $\bar\nu_\mu\rightarrow
\bar\nu_e$ channel and the E531 experiment at Fermilab \cite{E531} for the
$\nu_\mu\rightarrow \nu_\tau$ channel. Neither of these experiments shows
evidence
for neutrino oscillation on those channels. Their results are also presented as
exclusion areas in the two-neutrino oscillation  approximation in Fig.\
\ref{2family}. In terms of transition probabilities we find
\begin{equation}
\begin{array} {l}
\langle P_{e\mu}\rangle  \lesssim 1.5\times 10^{-3} \;\;\; \mbox{ from E776} \\
\langle P_{\mu\tau}\rangle   \lesssim 2\times 10^{-3}   \;\;\; \mbox{ from
E531} \\
\end{array}
\end{equation}
at $90 \%$ CL.

Recently the Liquid Scintillator Neutrino Detector (LSND) experiment
\cite{LSND} has announced the observation of an anomaly that can be
interpreted as  neutrino oscillations in the channel $\bar{\nu_\mu} \rightarrow
\bar{\nu_e}$. Most of the oscillation parameters required as explanation are
already excluded by the E776\cite{E776} and KARMEN \cite{karmen} experiments.
Their result can be marginally compatible with these previous results  for
$\Delta m^2 \approx 6$--$8$ eV$^2$ and mixing $\sin^2 (2\theta)\approx  3\times
10^{-3}$.  One must point out, however, that these ranges are still quite far
from certain.

\subsection{Solar Neutrinos}
At the moment, evidence for a solar neutrino deficit comes from four
experiments \cite{solar}. Putting all these results together seems to
indicate that the solution to the problem is not astrophysical but must
concern neutrino properties. The standard explanation for this deficit
would be the oscillation of $\nu_e$ to another neutrino species either active
or sterile. Different analyses have been performed to find the allowed mass
differences and mixing angles in the two-flavour approximation \cite{sol2fam}.
They all seem to agree that there are three possible solutions for the
parameters:\\
\noindent
$\bullet$ vacuum oscillations with $\Delta m^2_{ei}=(0.5$--$1)\times 10^{-10}$
eV$^2$ and $\sin^2(2\theta)=0.8$--$1$ \\
\noindent
$\bullet$ non-adiabatic-matter-enhanced oscillations via the MSW mechanism
\cite{msw}
with $ \Delta m^2_{ei}=(0.3$--$1.2)\times 10^{-5}$ eV$^2$ and
$\sin^2(2\theta)=(0.4$--$1.2)\times 10^{-2} $, and \\
\noindent
$\bullet$ large mixing via the MSW mechanism with
$\Delta m^2_{ei}=(0.3$--$3)\times 10^{-5}$  eV$^2$ and
$\sin^2(2\theta)=0.6$--$1$.

\subsection{Atmospheric Neutrinos}
Atmospheric neutrinos are produced when cosmic rays (primarily protons) hit the
atmosphere and initiate atmospheric cascades. The mesons present in the
cascade, decay leading to a flux of $\nu_e$ and $\nu_\mu$ which reach the Earth
and
interact in the different neutrino detectors. Naively the expected ratio of
$\nu_\mu$ to $\nu_e$ is in the proportion $2:1$, since
the main reaction is $\pi \rightarrow \mu \nu_\mu$ followed by
$\mu \rightarrow e \nu_\mu \nu_e$. However, the expected ratio
of muon-like interactions to electron-like interactions in each experiment
depends on the detector thresholds and efficiencies  as well as on the expected
neutrino fluxes.

Currently four experiments have observed atmospheric neutrino interactions.
Two experiments use water-Cherenkov detectors, Kamiokande
\cite{kamisub,kamimul} and IMB \cite{IMB}, and have observed  a ratio
of $\nu_\mu$-induced events to $\nu_e$-induced events  smaller than the
expected one. In particular Kamiokande has performed two different analyses for
both sub-GeV neutrinos \cite{kamisub} and multi-GeV neutrinos \cite{kamimul},
which show the same deficit. On the other hand, the results from
the two iron calorimeter
experiments, Fr\'ejus \cite{frejus} and NUSEX \cite{nusex}, appear to be in
agreement with the predictions.

The results of the three most precise experiments are shown next.
They are given as
a double ratio  $R_{\mu/e}/R^{MC}_{\mu/e}$ of experimental-to-expected ratio of
muon-like to electron-like events. The expected ratio $R^{MC}_{\mu/e}$ is
obtained by Monte Carlo calculation of the atmospheric neutrino fluxes. We
have used the expected fluxes from Gaisser {\sl et al.} \cite{gaisser}
(see also Ref.\ \cite{fogli}) and Volkova \cite{volkova} depending on the
neutrino energies
(see discussion in Sec.\ \ref{4famil}). Use of other flux calculations
\cite{otherflux} would yield similar numbers.
\begin{equation}
\begin{array}{l}
R_{\mu/e}/R^{MC}_{\mu/e}=0.55 \pm 0.11 \;\;\;\;\mbox{for IMB}\\
R_{\mu/e}/R^{MC}_{\mu/e}=0.6 \pm 0.09 \;\;\;\;\mbox{for Kamiokande sub-GeV}\\
R_{\mu/e}/R^{MC}_{\mu/e}=0.59 \pm 0.1 \;\;\;\;\mbox{for Kamiokande multi-GeV}\\
R_{\mu/e}/R^{MC}_{\mu/e}=1.06 \pm 0.23 \;\;\;\;\mbox{for Fr\'ejus}
\end{array}
\label{atmosdat}
\end{equation}
The most plausible explanation for this anomaly is to suppose that $\nu_\mu$
oscillates into another flavour. The oscillation $\nu_\mu\rightarrow \nu_e$
is almost completely ruled out by the reactor experiment data \cite{bugey}.
We are then left with $\nu_\mu\rightarrow  \nu_\tau$ oscillations or
oscillations to a sterile neutrino. The allowed range of masses and mixings
in the two-family approximation from a global fit to the previous data
is shown in Fig.\ \ref{atmos1}.a and can be summarized as
\begin{equation}
\Delta m^2_{\mu i}\gtrsim 2\times 10^{-3} \;\;\;\;
\sin^2(2\theta)=0.56\mbox{--}1\;.
\end{equation}

The experiments have also observed an angular dependence on the value of these
double ratios. The previous ranges were obtained without  using the angular
information. An analysis based on the angular  dependence of the ratio  leads
to
an upper limit on the possible value  of $\Delta m^2$, as can be seen from the
Kamiokande multi-GeV analysis \cite{kamimul} and the IMB data \cite{IMB}.
However the allowed regions obtained from the best fit to these two sets of
data are inconsistent at the 2$\sigma$ level (see Fig.\ 5 in
Refs.\ \cite{kamimul,at2fam}). For this reason, we have
chosen not to use the constraints arising from the angular dependence of
the data in this analysis.

\subsection{Dark Matter}
There is increasing evidence that more than $90\%$ of the mass in the Universe
is dark and non-baryonic. Neutrinos, if massive,  constitute a source for dark
matter. Stable neutrinos can fill the Universe of hot dark matter if their
masses add up to a maximum of about 30 eV.  However, scenarios with only hot
dark
matter run into trouble in the explanation of the formation of structures on
small
scales of the Universe.  On the other hand, models for structure formation
favour the presence of cold dark matter. These models, however,  fail in
reproducing the data on the  anisotropy of the microwave background.
Currently, the best scenario to explain the data considers a mixture of
70$\%$ cold plus 30$\%$ hot dark matter \cite{dark1}.
This  translates into an upper limit on neutrino
masses \cite{dark2}:
\begin{equation}
\sum_i m_{\nu_i}\lesssim 4\mbox{--}7\; \mbox{eV}\; .
\end{equation}

\section{Four-Flavour Models}
\label{4famil}
Naive two-family counting shows that it is very difficult to  fit all
experimental information in Sec.\ \ref{exp} with three neutrino flavours
\cite{valle1,moha}, even without invoking the LSND data.  The solar neutrino
deficit could be due to $\nu_e\rightarrow \nu_\mu$  oscillations and the
atmospheric
neutrino deficit to $\nu_\mu\rightarrow \nu_\tau$ oscillations with the
appropriate mass differences, for example with a mass hierarchy
$m_{\nu_\tau}\gg m_{\nu_\mu}, m_{\nu_e}$. However, fitting this together with
the present laboratory limits leaves no room for hot dark matter since the
maximum allowed mass is of about $m_{\nu_\tau} < 0.7$ eV \cite{fogli}. The only
possible way out is to require that all three neutrinos are almost degenerate.
This requires a certain degree of fine-tuning in order to explain the
neutrinoless double beta decay data \cite{moha,inverted}. \footnote{Notice that
this scenario is also inconsistent with the  oscillation parameters observed by
LSND, should this experiment confirm their results}

One could have $\nu_\mu\rightarrow \nu_\tau$ oscillations for the atmospheric
neutrino deficit with  almost degenerate $\nu_\mu$ and $\nu_\tau$ with masses
$m_{\nu_\mu}=m_{\nu_\tau}\approx 2.5$ eV and $m_{\nu_e}\approx 0$
\cite{moha,new},  but leaving out the explanation for the solar neutrino
deficit.
Or $m_{\nu_\mu}=m_{\nu_\tau}\approx 0$ eV and $m_{\nu_e}\approx 2.5$ to explain
the atmospheric data but leaving unexplained  both solar neutrino deficit and
dark matter \cite{new}.

Also, it is possible to explain the solar neutrino deficit with
$\nu_e\rightarrow \nu_{\tau(\mu)}$ with almost degenerate $\nu_e$ and
$\nu_{\tau(\mu)}$ with masses   $m_{\nu_e}=m_{\nu_{\tau(\mu)}}\approx 2.5$ eV
and $m_{\nu_{\mu (\tau)}}\approx 0$, but leaving the atmospheric neutrino
deficit  unexplained \cite{inverted,invermoha}. Also
$m_{\nu_e}=m_{\nu_{\tau(\mu)}}\approx 0$ eV and $m_{\nu_{\mu (\tau)}}
\approx 2.5$ eV would explain the LSND data if confirmed but leaves  both
atmospheric and dark matter without explanation.

In the spirit of Pauli, one is tempted to introduce a
new neutrino as a ``desperate solution'' \cite{pauli} to
understand all present data. The nature of such a particle is
constrained by  LEP results on the invisible $Z$ width as well as data on the
primordial $^4$He abundance. Those rule out the existence of additional,
light, active neutrinos. In consequence the fourth neutrino state must be
sterile.

There are different mass patterns that one can construct with four such
neutrinos to verify all the experimental constraints and evidence for neutrino
oscillation and masses. We will assume a {\sl natural} mass hierarchy  with two
light neutrinos with their main projection in the $\nu_s$ and $\nu_e$
directions
and two heavy neutrinos with their largest component along the $\nu_\mu$ and
$\nu_\tau$ flavours. We will also require that the sterile neutrino does not
mix directly with the two heavy states. As we will see, this is necessary to
verify the constraints from big bang  nucleosynthesis \cite{BBN}.  Such a
hierarchy appears naturally, for instance if one advocates an $L_e\pm L_\mu\mp
L_\tau$ discrete symmetry for the mass matrix \cite{valle1}. In Ref.\
\cite{moha} a similar mass pattern is also generated via a combination of
see-saw mechanism and loop mechanism. Other possible patterns which explain
all data such as some inverted mass schemes \cite{inverted} require a somehow
large degree of fine-tuning to explain the absence of neutrinoless double beta
decay.

For any matrix presenting this {\sl natural} mass hierarchy, the mixing matrix
can be parametrized in a general way as \footnote{A similar mass matrix is
also introduced in Ref. \cite{valle2}.}
\begin{equation}
U=\begin{array}{c|cccc}
 & \nu_s & \nu_e & \nu_\mu & \nu_\tau \\
\hline
\nu_1 & c_{es} & s_{es}c_{e\mu}c_{e\tau} & s_{es}s_{e\mu}
& s_{es}c_{e\mu}s_{e\tau} \\
\nu_2 & -s_{es} & c_{es}c_{e\mu}c_{e\tau} & c_{es} s_{e\mu}
& c_{es}c_{e\mu} s_{e\tau} \\
\nu_3 & 0 & -c_{\mu\tau}s_{e\mu}c_{e\tau}-s_{\mu\tau}s_{e\tau}
& c_{\mu\tau}c_{e\mu} & -c_{\mu\tau}s_{e\mu}s_{e\tau}+s_{\mu\tau}c_{e\tau} \\
\nu_4 & 0 & s_{\mu\tau}s_{e\mu}c_{e\tau}-c_{\mu\tau}s_{e\tau}
& -s_{\mu\tau}c_{e\mu} & s_{\mu\tau}s_{e\mu}s_{e\tau}+c_{\mu\tau}c_{e\tau} \\
\end{array}
\end{equation}
with $c_i=\cos\theta_i$ and $s_i=\sin\theta_i$.
For the sake of simplicity we have assumed no CP violation
in the lepton sector.
In this approximation  $m_1,m_2\ll m_3,m_4$ and
$|m_3^2-m_4^2| \ll m_3^2,m_4^2$. Also, $\nu_3$ and $\nu_4$ constitute 30$\%$ of
dark matter in the Universe. This implies $m_3\simeq m_4=2$--$3.5$ eV. Such
a mass pattern has been argued to yield satisfactory results in Cold+Hot
Dark Matter scenarios \cite{dark2}.
We can define
\begin{equation}
\begin{array}{l}
\Delta m^2_{solar}= |m_1^2-m_2^2| \\
\Delta m^2_{AT}=|m_3^2-m_4^2| \\
\Delta M^2_{DM}=|m_1^2-m_3^2|\simeq |m_1^2-m_4^2|\simeq |m_2^2-m_3^2|
\simeq |m_2^2-m_4^2|\simeq (4-10)\; \mbox{eV}^2\; . \\
\end{array}
\end{equation}

Transition probabilities between the different flavours will now have
contributions from the three oscillation lengths due to the three different
mass differences in the problem which we will denote
$\sin^2({\Delta_{solar}}/{2})$,  $\sin^2({\Delta_{AT}}/{2})$, and
$\sin^2({\Delta_{DM}}/{2})$, respectively.  We are now ready to reanalyse the
experimental data presented in Sec.\ \ref{exp} in the four-flavour framework
considering the oscillations with  the three different oscillation lengths.

\section{Global Analysis}
\label{global}
\subsection{Laboratory Experiments}
Laboratory experiments are insensitive to oscillations due to the solar mass
difference $\Delta m^2_{solar}$. Also, for most of them the oscillation
due to the dark matter mass difference will be averaged to $1/2$. As a
consequence the negative results presented in Sec.\ \ref{exp} will impose
severe constraints on the mixing angles.

For the Bugey reactor experiment the relevant transition probability is the
$\nu_e$ survival probability. For any value of the atmospheric mass difference
this probability will always verify
\begin{equation}
0.93 \lesssim P_{ee}^{Bugey} \leq  1-
2 c_{e\mu}^2 c_{e\tau}^2(1- c_{e\mu}^2 c_{e\tau}^2)
\;\; \Rightarrow \;\;  c_{e\mu}^2 c_{e\tau}^2\geq 0.96\; .
\label{bugey}
\end{equation}
For CDHSW the relevant probability is the $\nu_\mu$ survival probability
\begin{equation}
0.95\lesssim P_{\mu\mu}^{CDHSW}\leq 1- \frac{1}{2}\sin^2(2 \theta_{e\mu})
\;\; \Rightarrow \;\;  \sin^2(2 \theta_{e\mu})\lesssim 0.1\; .
\end{equation}
For E776 the situation is somehow more involved,  since the value of the
oscillating phase $\langle \sin^2( {\Delta_{DM}}/{2})\rangle $ varies in the
range
$\Delta_{DM}=4$--$10$ eV$^2$ due to the wiggles of the resolution function of
the
experiment (see Fig.\ \ref{2family}). Also, the experiment is sensitive to the
atmospheric mass difference:
\begin{equation}
\begin{array}{ll}
1.5\times 10^{-3}\geq  & P_{e\mu}^{E776}=
 \sin^2(2\theta_{e\mu}) c^2_{e\tau} \sin^2(\frac{\Delta_{DM}}{2}) \\
& +\,[\frac{1}{2} \sin(2\theta_{\mu\tau})\sin(2\theta_{e\mu})
\sin(2\theta_{e\tau})\cos(2\theta_{\mu\tau})c_{e\mu}-\sin^2(2\theta_{\mu\tau})
(s^2_{e\mu}c^2_{e\tau}-s^2_{e\tau})c^2_{e\mu}]\sin^2(\frac{\Delta_{AT}}{2})\; .
\end{array}
\end{equation}
For any value of the atmospheric mass difference and the $\mu\tau$ mixing
angle, the previous limit is verified if
\begin{equation}
\sin^2(2\theta_{e\mu}) c^2_{e\tau}\leq (2\mbox{--}5)\times 10^{-3}
\end{equation}
The limit from E531 on the mixings $e\mu$ and $e\tau$  is always less
restrictive than the previous ones for any value of $\Delta m^2_{AT}$ and
$\Delta M^2_{DM}$.

Combining these constraints we obtain that $e\mu$ and $e\tau$ mixings are
constrained to
\begin{equation}
\begin{array}{l}
\sin^2 (2\theta_{e\mu}) \leq (2\mbox{--}5)\times 10^{-3} \\
\sin^2 (2\theta_{e\tau}) \leq 0.16\;\;\;\;
\end{array}
\label{acc}
\end{equation}
where the range of $\sin^2 (2\theta_{e\mu})$ depends on the specific value
of $\Delta M^2_{DM}$.

If we now turn to the effect due to the oscillation with $\Delta_{AT}$,
we can rewrite the relevant probabilities for the different experiments
expanding in the small angles $e\mu$ and $e\tau$:
\begin{equation}
\begin{array}{l}
P_{\mu\mu}^{CDHSW}\simeq
1-\frac{1}{2}\sin^2(2\theta_{e\mu})-\sin^2(2\theta_{\mu\tau})
\sin^2(\frac{\Delta_{AT}}{2})\\
P_{\mu\tau}^{E531}\simeq \sin^2(2\theta_{\mu\tau}) c^2_{e\tau}\sin^2(\frac
{\Delta_{AT}}{2})\\
P_{e\mu}^{776}\simeq \sin^2(2\theta_{e\mu})c^2_{e\tau}
\sin^2(\frac{\Delta_{DM}}{2})
+ \sin^2(2\theta_{\mu\tau}) s^2_{e\tau} \sin^2(\frac{\Delta_{AT}}{2}) \; .\\
\end{array}
\end{equation}
With the constraints in Eq.\ (\ref{acc}), the Bugey experiment is not sensitive
to oscillations with $\Delta_{AT}$.

In Figs.\ \ref{atmos1} and \ref{atmos2} we show the exclusion contours in
($\Delta m^2_{AT}, \sin^2(2\theta_{\mu\tau})$) which are due to the three
experiments for different allowed values of the $e\mu$ and $e\tau$ mixings.

\subsection{Solar neutrinos}
Different authors have considered the propagation of neutrinos in the framework
of more-than-two-neutrino oscillations \cite{fogli,sol3fam,kuo,other3}.  Rather
than reanalysing the solar neutrino data we will follow and adapt the results
in Ref.\ \cite{fogli}.  Following the approach of Ref.\ \cite{kuo} we
will express the transition probabilities in the framework of four neutrinos in
terms of the two-family ones.

For the solar neutrino deficit the relevant transition probability is the
$\nu_e$ survival probability. For solar neutrino distance and energies,
both oscillations with the $\Delta m^2_{AT}$ and $\Delta M^2_{DM}$
will have averaged to $1/2$ and the survival probability in vacuum takes
the form
\begin{equation}
P_{ee}^{solar}\simeq
1-2 c_{e\mu}^2 c_{e\tau}^2(1- c_{e\mu}^2 c_{e\tau}^2) -
c_{e\mu}^2 c_{e\tau}^2\sin^2(2\theta_{es})
\langle \sin^2 (\frac{\Delta_{solar}}{2})\rangle \; ,
\end{equation}
where we have expanded in the small mixings $e\mu$ and $e\tau$. Given the
constraints on the product $c_{e\mu}^2 c_{e\tau}^2$ (Eq.\ (\ref{bugey})), the
effect
of the new mixings is very small and the vacuum oscillation solution remains
basically unchanged.

For propagation in matter we follow the notation given in Ref.\ \cite{kuo}.
Following a similar procedure we can write the transition probability in the
four-neutrino scenario in terms of the transition probability for two neutrinos
as
\begin{equation}
P_{ee}^{4\nu}= 1-c_{e\mu}^2 c_{e\tau}^2+c_{e\mu}^2 c_{e\tau}^2 P_{ee}^{2\nu}\;
,
\end{equation}
where $P_{ee}^{2\nu}$ is computed in the two-family case from the evolution
equation in matter substituting the electron density  $N_e$ by   $N_e
c_{e\mu}^2 c_{e\tau}^2$. According to the results of Ref.\ \cite{fogli},   and
taking into account Eq.\ (\ref{bugey}), the two MSW solutions are still valid
in the presence of the new mixings. The effect of the mixings appears to go in
the direction of favouring the small mixing solution \cite{fogli}.  The large
mixing solution is also in conflict with the constraints from  big bang
nucleosynthesis. As can be seen from Fig.\ 1 and Fig.\ 2 in Ref.\  \cite{BBN}
for $\nu_e$--$\nu_s$ oscillations this solution would lead to an  excess on the
primordial $^4$He abundance excluded by the present data.

\subsection{Atmospheric Neutrinos}
There are in the literature several analyses of the atmospheric neutrino data
in terms of neutrino oscillations in two-family and three-family scenarios
\cite{at2fam,at3fam,fogli}, the last ones being mostly in the
one-mass-dominance approximation. This approximation is not valid in our scheme
since  the atmospheric neutrino fluxes can show the effect of oscillations due
to  two oscillation lengths $\Delta_{AT}$ and $\Delta_{DM}$. We reanalyse the
data in Eq.\ (\ref{atmosdat}), taking into account the effect of these two
oscillations. In doing so we will follow the notation of  Ref.\ \cite{at3fam}.

In each experiment the number of $\mu$ events, $N_\mu$, and of $e$ events,
$N_e$, in the presence of oscillations will be
\begin{equation}
N_\mu=N^0_{\mu\mu} \langle P_{\mu\mu}\rangle +N^0_{e\mu} \
\langle P_{e\mu}\rangle \; ,  \;\;\;\;\;\
N_e=N^0_{ee} \langle P_{ee}\rangle +N^0_{\mu e} \langle P_{\mu e}\rangle \; ,
\end{equation}
where
\begin{equation}
N^0_{\alpha\beta}=\int \frac{d^2\Phi_\alpha}{dE_\nu d\cos\theta_\nu}
\frac{d\sigma}{dE_\beta}\epsilon(E_\beta)
dE_\nu dE_\beta d(\cos\theta_\nu)
\end{equation}
and
\begin{equation}
\langle P_{\alpha\beta}\rangle =\frac{1}{N^0_{\alpha\beta}}\int
\frac{d^2\Phi_\alpha}{dE_\nu d\cos\theta_\nu} P_{\alpha\beta}
\frac{d\sigma}{dE_\beta}\epsilon(E_\beta)
dE_\nu dE_\beta d(\cos\theta_\nu)\; .
\end{equation}
Here $E_\nu$ is the neutrino energy and $\Phi_\alpha$ is the flux of
atmospheric
neutrinos $\nu_\alpha$; $E_\beta$ is the final charged lepton energy and
$\epsilon(E_\beta)$ is the detection efficiency for such charged lepton;
$\sigma$ is the interaction cross section $\nu \; N \rightarrow N'\; l$.
The expected rate with no oscillation would be
\begin{equation}
R^{MC}_{\mu/e}= \frac{N^0_{\mu\mu}}{N^0_{ee}}\; .
\end{equation}
The value of this ratio is different for each experiment as it depends on
the threshold for the detected lepton energy as well as on the detection
efficiency for the different lepton flavours.
The double ratio $R_{\mu/e}/R^{MC}_{\mu/e}$ of the expected ratio of muon-like
to electron-like events with oscillation to the expected ratio without
oscillations  is given by
\begin{equation}
\frac{R_{\mu/e}}{R^{MC}_{\mu/e}}=
\frac{\langle P_{\mu\mu}\rangle +\frac{N^0_{e\mu}}{N^0_{\mu\mu}}
\langle P_{e\mu}\rangle }
{\langle P_{ee}\rangle +\frac{N^0_{\mu e}}{N^0_{ee}}
\langle P_{\mu e}\rangle  } \; .
\label{atratio}
\end{equation}
We perform a global fit to the data in Eq.\ (\ref{atmosdat}) using the neutrino
fluxes from Ref.\ \cite{gaisser} for $E_\nu<3$ GeV. For  $E_\nu>10$ GeV we use
the fluxes from Ref.\ \cite{volkova}, and for energies in between the  two
fluxes
are smoothly connected. We perform the integrals starting at the different
thresholds of each experiment and include the published detection efficiencies
\cite{kamisub,kamimul,IMB,frejus}. For the neutrino interaction cross section
the dominant process is the quasielastic cross section \cite{smith}. For
larger neutrino energies (for Fr\'ejus and Kamiokande multi-GeV analyses)
one-pion and multipion cross sections become relevant  and are included
\cite{cross}.

\begin{figure}
\begin{center}
\mbox{\epsfig{file=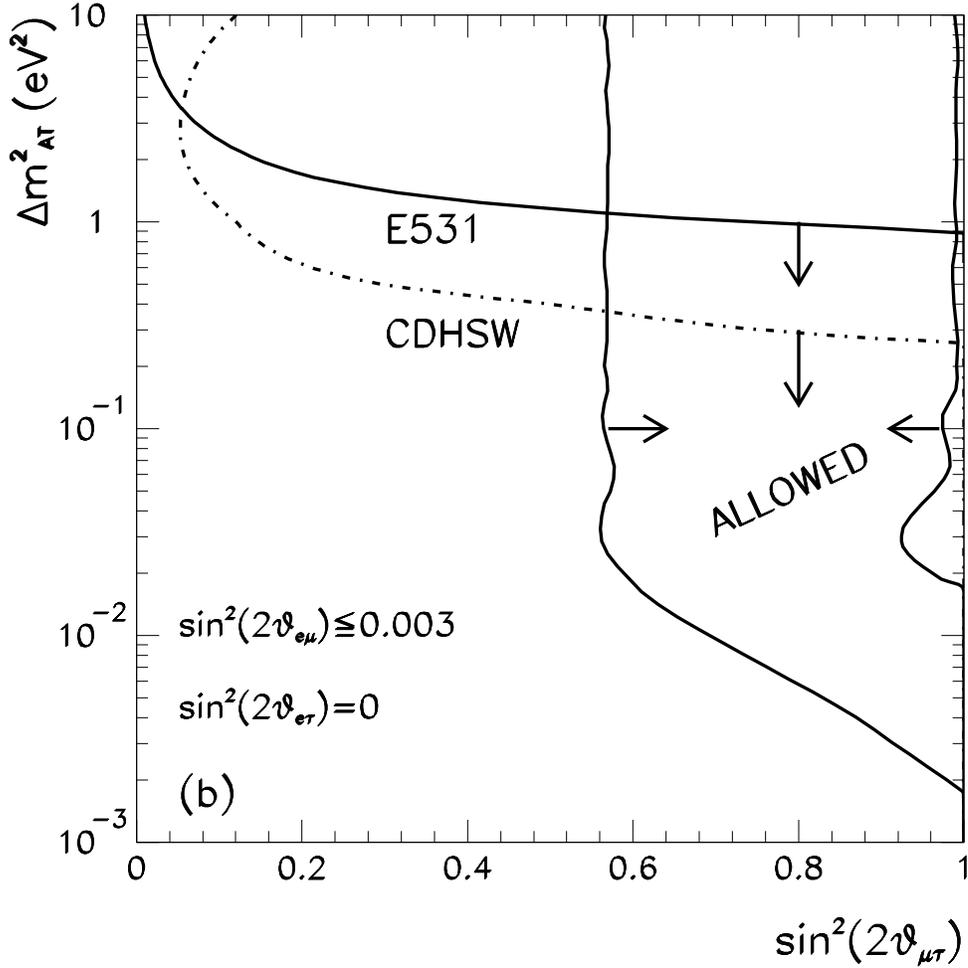,bbllx=60,bblly=200,bburx=600,bbury=600}}
\end{center}
\caption{{\bf (a)} The 2$\sigma$ allowed regions in the $\Delta m^2_{AT}$,
$\sin^2(2\theta_{\mu\tau})$ plane  for the atmospheric neutrino experiments for
zero mixing $e\tau$ and allowed values of mixing $e\mu$: Fr\'ejus (solid line),
Kamiokande sub-GeV events (dashed line), IMB (dotted line), and Kamiokande
multi-GeV events (dot-dashed line). The arrows point towards the allowed
regions for the different experiments.  The thick solid line surrounds the
2$\sigma$ allowed region for the combination of all experiments. \\
{\bf (b)}.  Allowed region from the atmospheric neutrino deficit analysis
together with the relevant constraints from the laboratory experiments
for zero mixing $e\tau$ and allowed values of mixing $e\mu$. }
\label{atmos1}
\end{figure}

\begin{figure}
\begin{center}
\mbox{\epsfig{file=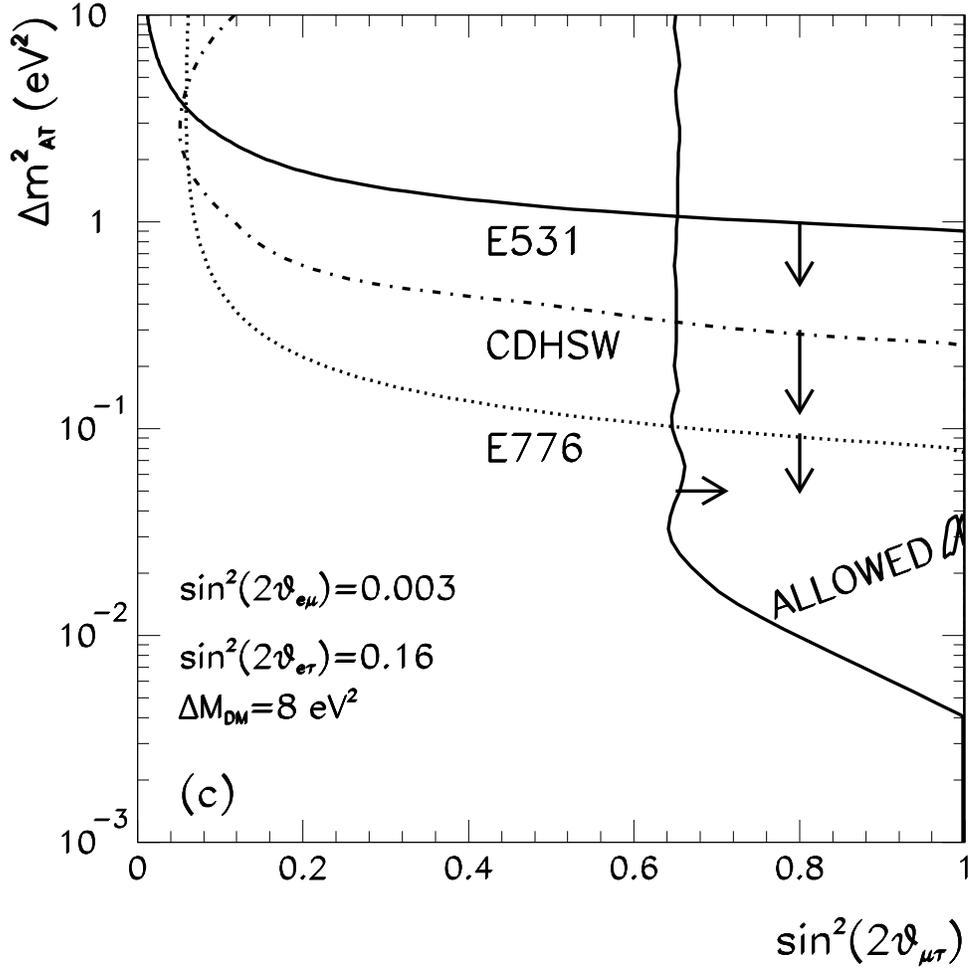,bbllx=60,bblly=200,bburx=600,bbury=600}}
\end{center}
\caption{{\bf (a)} Same as Fig. 2.a for
maximum allowed mixing $e\tau$ and allowed values of the mixing $e\mu$.\\
{\bf (b)} Same as Fig. 2.b for
maximum allowed mixing $e\tau$ and allowed values of the mixing $e\mu$. \\
{\bf (c)}.  Same as {\bf (b)} but for maximum allowed mixing $e\mu$ and
corresponding allowed value for $\Delta M^2_{DM}$ from the E776 experiment
data.}
\label{atmos2}
\end{figure}

The results of our $\chi^2$ fit are shown in Figs.\ \ref{atmos1} and
\ref{atmos2}.  In
Fig.\ \ref{atmos1} the results are shown for zero mixings $e\mu$ and $e\tau$ as
in a two-family scenario. Figure  \ref{atmos2} shows the effect of the
inclusion of
the mixings. As seen in the figure the inclusion of the $e\tau$ mixing leads
to a  more constrained area for the oscillation parameters.  When mixing $e\mu$
and $e\tau$ are non-zero and taking into account the  constraints in
Eq.\ (\ref{acc}), Eq.\ (\ref{atratio}) takes the form
\begin{equation}
\frac{R_{\mu/e}}{R^{MC}_{\mu/e}}\simeq
1+ \frac{1}{2}\sin^2(2\theta_{e\tau})
-\sin^2(2\theta_{\mu\tau}) \langle \sin^2(\frac{\Delta_{AT}}{2})\rangle \; ,
\label{ratmix}
\end{equation}
where we have used the fact that for atmospheric neutrino experiments
$\langle \sin^2({\Delta_{solar}}/{2})\rangle =0$ and $\langle
\sin^2({\Delta_{DM}}/{2})\rangle =1/2$.
For the purpose of illustration we have used the approximation
${N^0_{e\mu}}/{N^0_{\mu\mu}}={N^0_{ee}}/{N^0_{\mu e}}\simeq 0.5$.  The effect
of
the $e\tau$ mixing in Eq.\ (\ref{ratmix}) is to increase the value of the
double
ratio. Therefore a larger amount of $\mu\tau$ oscillation is needed to account
for the deficit. Due to the small values allowed, a non-zero mixing $e\mu$ does
not modify the  analysis of the atmospheric neutrino data provided that
($\Delta M^2_{DM} , \sin^2(2\theta_{e\mu}$)) are allowed by the E776 data (see
Fig.\ \ref{2family}).

\section {$\nu_\tau$ Oscillation Experiments}
\label{experiments}
\subsection{CHORUS and NOMAD}
\begin{figure}
\begin{center}
\mbox{\epsfig{file=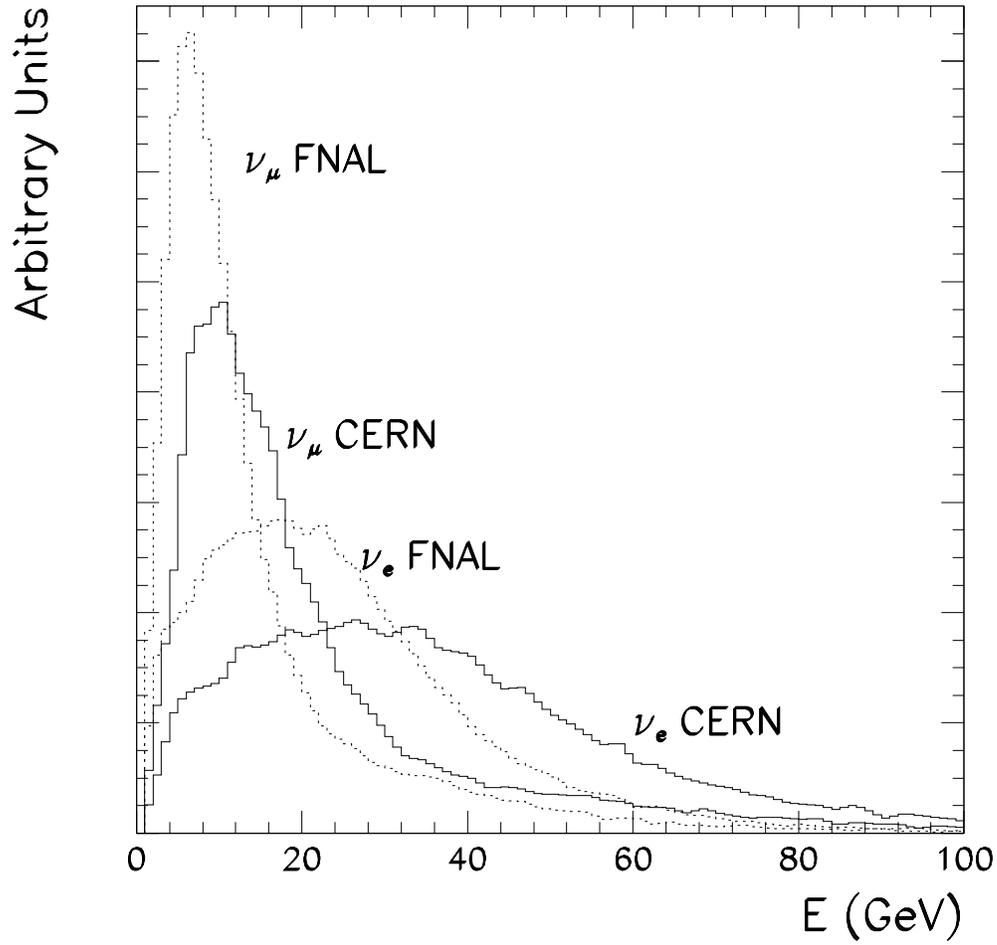,bbllx=60,bblly=200,bburx=600,bbury=600}}
\end{center}
\caption{$\nu_e$ and $\nu_\mu$ spectra for the CERN beam (solid lines)
and the Fermilab beam (dashed lines) as labeled.}
\label{fluxes}
\end{figure}

The two upcoming $\nu_\mu(\nu_e) \rightarrow \nu_\tau$ experiments, CHORUS and
NOMAD, are $\nu_\tau$ appearance experiments, i.e., they search for the
appearance of $\nu_\tau$'s  in  the CERN SPS beam consisting primarily of
$\nu_\mu$'s, with about 1\% $\nu_e$'s, as shown in  Fig.\
\ref{fluxes} (solid lines) \cite{chorus,nomad}.
The mean energy of the $\nu_\mu$ beam is around
$30$ GeV and the detectors are located approximately $800$ m away from the beam
source. The  $\nu_\tau$ contamination of the SPS  beam is virtually zero.

CHORUS and NOMAD have complementary techniques to identify a $\tau$ lepton. The
NOMAD experiment \cite{nomad} distinguishes a $\nu_\tau$ CC
interaction from ordinary $\nu_\mu$ or $\nu_e$ interactions by exploiting
the fact that the $\tau$ lepton produced in a $\nu_\tau$ CC interaction, decays
emitting one
or more neutrinos which result in a measurable amount  of missing transverse
momentum, in the  general direction of the charged lepton. NOMAD is in essence
an electronic bubble chamber, with a continuous target alternating panels of a
light material and drift chambers, followed by  a
transition radiation detector and an electromagnetic calorimeter. All the
above detectors are located inside a $0.4$ T magnetic field. Thus NOMAD
measures essentially all the charged tracks and photons in the event, which
enables a good reconstruction of the transverse missing momentum magnitude and
direction. CHORUS \cite{chorus} on the other hand, seeks to observe the finite
path of the $\tau$ on its emulsion target. At SPS energies, the $\tau$ mean
decay length is about  $1$ mm, giving two distinctive signatures in CHORUS
emulsion.   One-prong decays result in a short track with a kink, while
three-prong decays appear as a short track splitting in three. Thus CHORUS rely
on purely vertex techniques to identify the $\tau$. However, in order to reduce
the number of events to be scanned, loose kinematical cuts  are also applied.
To this extent, the emulsion target is followed by a spectrometer and a
compensating calorimeter. In the surviving events, tracks reconstructed in the
spectrometer are extrapolated to the emulsions in order to determine where to
scan.

The initial goal of the SPS is to deliver $2.4 \times 10^{19}$ protons on
target over two years to both experiments. The intense neutrino beam allows
very light detectors (NOMAD has a fiducial mass of near 3 t, while the
emulsion target of CHORUS is 800 kg). The main characteristics of NOMAD and
CHORUS as well as their expected performance are summarized in Table
\ref{tab:nomad}.
\subsection{Future $\nu_\tau$ Experiments at CERN}
There are a number of future $\nu_\mu(\nu_e) \rightarrow \nu_\tau$ experiments
beyond CHORUS and NOMAD being discussed at present. In the more immediate
future there exists the possibility of an extended run at the CERN SPS beam,
spanning several years after the initial period. Several suggestions have been
made to upgrade CHORUS and NOMAD. The upgrade proposed for CHORUS would go in
the direction of substituting the emulsion target with an active target made of
scintillating capillaries read-out by CCDS \cite{capilar}. NOMAD, on the other
hand, could be upgraded by adding an instrumented target made of a sandwich of
passive, low-$Z$ material (carbon or beryllium) and silicon detectors
\cite{nomad01}. Ultimately, one would like a precise measurement of the event
vertex, which,  when combined with the event kinematics, results in a  much
improved sensitivity. Specifically, the sensitivity of the upgraded NOMAD
detector could be improved by one order of  magnitude \cite{nomad01}. A
different experimental approach, suggested in \cite{carlo}, proposes a 100 t
liquid-CH$_4$ TPC in a high magnetic field (2 t). The idea is to use the
quasi-free protons provided by the CH$_4$ to completely constrain the
kinematics of the  $\nu_\tau$--proton collisions. One can determine the
momentum of the outgoing
neutrino in the case of a $\tau$ decay involving one neutrino, and reconstruct
invariant masses. The $\tau$ signal would appear as a mass peak over a very low
background reduced by cuts \'a la CHORUS/NOMAD.

As a specific example of this improved detector we have considered
a suggestion to upgrade the NOMAD detector (see \cite{nomad01} where detailed
calculations are carried out). We will refer to this future detector with
the generic name of Neutrino ApparatUS with Improved CApAbilities (NAUSICAA).
Such a device would permit $\tau$ detection using $both$ kinematical and
vertex techniques. Neutrino interactions would occur in a target made of a
sandwich of silicon detector and a light-$Z$ material.
The event vertex would be precisely measured using the silicon
information. The target would be followed by several planes of drift
chambers and a calorimeter, with the full detector inside a
magnetic field of around 1 T. The detector performance is summarized in
Table \ref{tab:nausicaa}.
NAUSICAA can effectively suppress the CC and Neutral Currents (NC)
$\nu_\mu (\nu_e)$ backgrounds, thanks to the combination of two independent
signatures of the
$\tau$, i.e., its ``long'' lifetime and its decays with neutrino emission.

Finally, there are several proposals for long-baseline experiments using the
CERN SPS beam (see for example \cite{revol} and references therein).
\subsection{Future $\nu_\tau$ Experiments at FNAL}
At Fermilab, a neutrino beam will be available when the main injector becomes
operational, around the year 2001. Compared with the CERN SPS beam, the  main
injector will deliver a beam 50 times more intense, but with an average energy
around one third of that of the SPS neutrinos \cite{E803,minos}
(see Fig.\ \ref{fluxes}).

There are currently two experiments proposed to operate in this beam. One is a
short-baseline experiment, E803 \cite{E803}, very similar in design and
fiducial mass to CHORUS. Nevertheless, E803 is foreseen to have a sensitivity
ten times
better than CHORUS, thanks essentially to a  much larger data sample  (in
addition to a more intense beam, the experiment expects a 4-year run). MINOS
\cite{minos} is a long-baseline experiment, which proposes two detectors,
separated by 732 km. This experiment can perform several tests to look for a
possible oscillation $\nu_\mu \rightarrow \nu_\tau$ in the small mass
difference range. Of those, the most sensitive is the comparison of the
fraction of CC-like (defined for $\nu_\mu$) and NC-like
events (a $\nu_\tau$ event, appears, except for the decay  $\tau \rightarrow
\mu \nu_\mu \nu_\tau$ as a neutral current in the Minos detector).
The main characteristics and expected performance of E803 and MINOS are
summarized in Table \ref{tab:e803}.

Finally we will consider the possibility of installing the NAUSICAA detector
(see Table \ref{tab:nausicaa})  as an alternative or a successor to E803 in the
Fermilab beam.

\section{Discovery Potential}
\label{discovery}
We now turn to the study of the prospects for the discovery of $\nu_\tau$
oscillations
in the upcoming CERN experiments CHORUS and NOMAD, in the future
experiment E803 at Fermilab as well as a possible improved experiment,
NAUSICAA,
operated both at CERN and Fermilab beam. As an example of long-baseline
experiment we also analyse the prospects for MINOS.

\begin{figure}
\begin{center}
\mbox{\epsfig{file=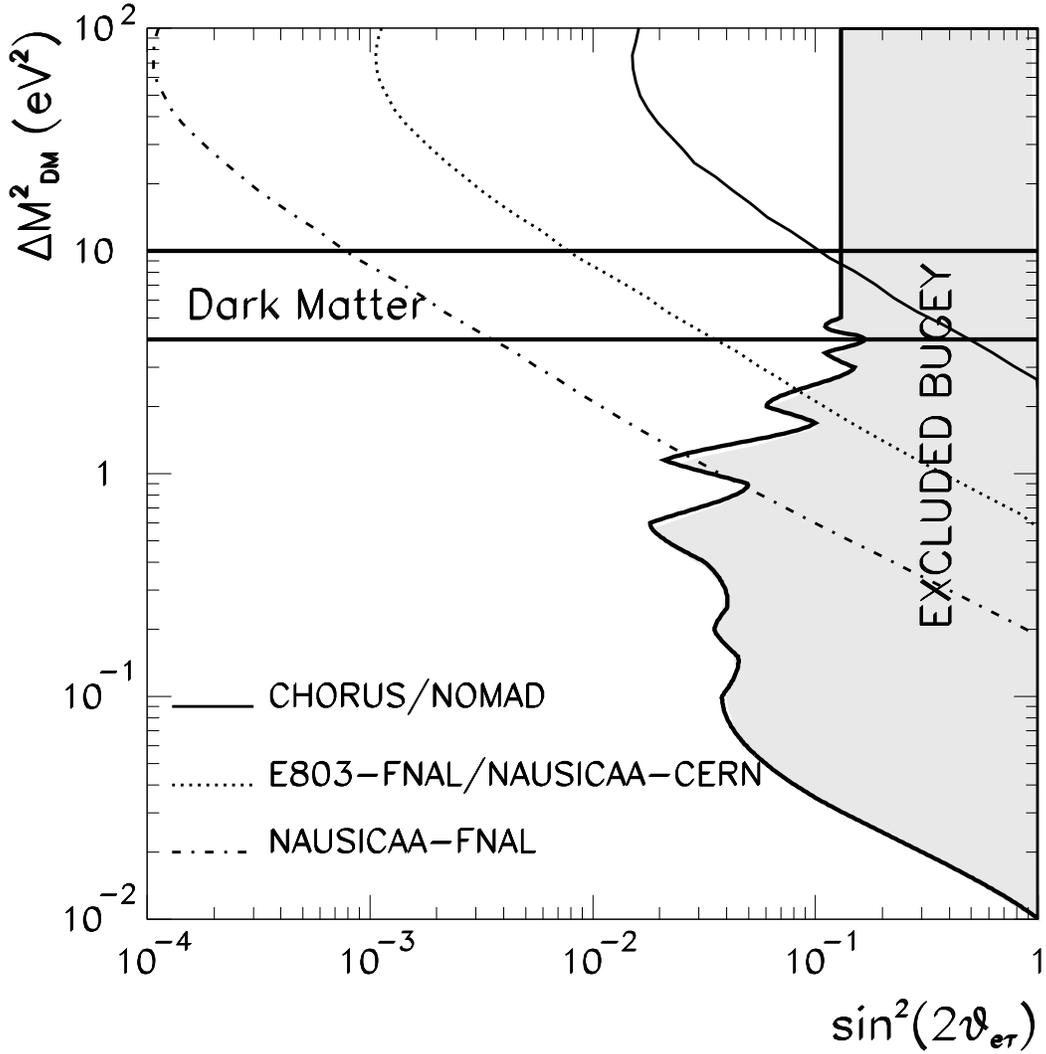,bbllx=60,bblly=300,bburx=600,bbury=700}}
\end{center}
\caption{Accessible regions (90\% CL) for the $\nu_e\rightarrow \nu_\tau$
oscillation in the ($\Delta M^2_{DM}$, $\sin^2(2\theta_{e\tau})$) plane for the
CHORUS/NOMAD  experiments (fine solid line) and Fermilab experiment  E803/the
improved CERN experiment  (dotted line).  The dot-dashed line delimits the
region accessible to NAUSICAA at Fermilab.  Also shown in the figure are the
region at  present excluded  by the Bugey  data (shaded area) and the favoured
range on $\Delta M^2_{DM}$ from dark matter considerations (solid horizontal
lines). }
\label{etau}
\end{figure}

After implementing the limits derived in Sec.\ \ref{global} and considering the
sensitivity of the experiments, one sees that for all facilities the only
observable $\nu_e\rightarrow \nu_\tau$ transition oscillates with an
oscillation length $\Delta_{DM}$ such that
\begin{equation}
 P_{e\tau} \simeq \sin^2(2\theta_{e\tau})\sin^2(\frac{\Delta_{DM}}{2})
\label{et}
\end{equation}
Figure \ref{etau} shows the regions accessible to the experiments in  the
$(\sin^2(2\theta_{e\tau}),\Delta M^2_{DM})$ plane. As can be seen, the almost
complete accessible range of parameters  for the CERN experiments CHORUS and
NOMAD is already excluded by the Bugey data in the favoured range for the mass
difference due to dark matter considerations.  E803 at Fermilab and NAUSICAA at
CERN could observe these oscillations in the whole favoured range of $\Delta
M^2_{DM}$, provided  that  $\sin^2(2\theta_{e\tau})\gtrsim 3\times 10^{-2}$.
Furthermore, NAUSICAA operated at the Fermilab beam  can go as low as
$\sin^2(2\theta_{e\tau}) \gtrsim 4\times 10^{-3}$. MINOS, being a $\nu_\mu$
disappearance experiment is not sensitive to this oscillation.

For transitions $\nu_\mu\rightarrow \nu_\tau$ a four-neutrino framework
predicts (unlike the naive two-family framework) $two$ oscillations, dominated
by the characteristic lengths  $\Delta_{DM}$ and $\Delta_{AT}$. All experiments
are in principle sensitive to both oscillations depending on the values of the
mixing angles
\begin{equation}
\begin{array}{l}
P^{DM}_{\mu\tau} \simeq \sin^2(2\theta_{e\mu})\sin^2(\theta_{e\tau})
\sin^2(\frac{\Delta_{DM}}{2}) \\
P^{AT}_{\mu\tau} \simeq \sin^2(2\theta_{\mu\tau})\cos^2(\theta_{e\tau})
\sin^2(\frac{\Delta_{AT}}{2}) \\
\end{array}
\label{mt}
\end{equation}

\begin{figure}
\begin{center}
\mbox{\epsfig{file=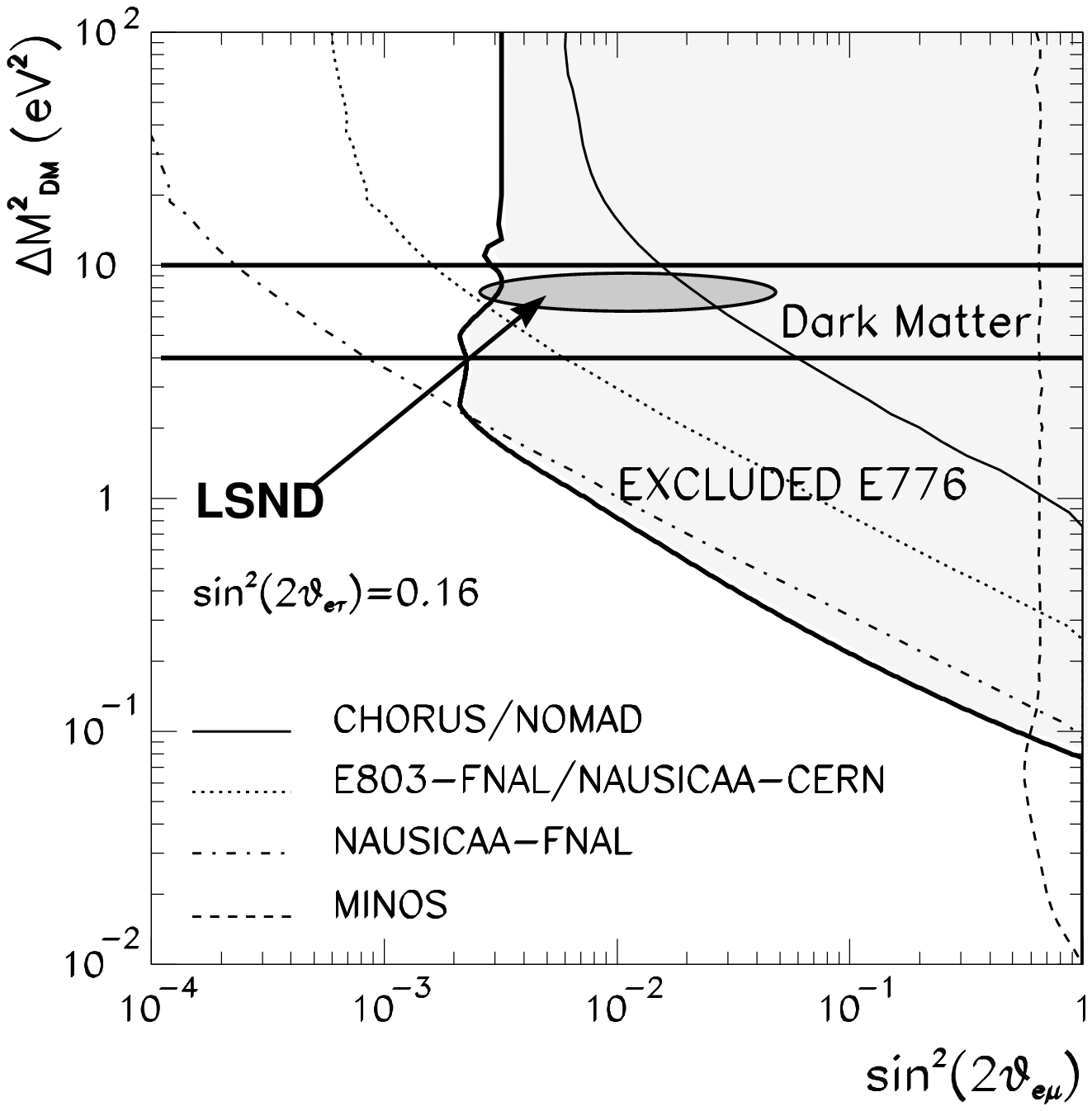,bbllx=60,bblly=300,bburx=600,bbury=700}}
\end{center}
\caption{Accessible regions (90\% CL) for the $\nu_\mu\rightarrow \nu_\tau$
oscillation in the ($\Delta M^2_{DM}$, $\sin^2(2\theta_{e\mu})$) plane  for an
optimum value of the mixing $e\tau$. The fine solid line corresponds  to the
CHORUS/NOMAD experiments and the dotted line corresponds to the Fermilab
experiment E803 and to the improved CERN experiment. The dot-dashed line
corresponds to NAUSICAA at Fermilab and the dashed line corresponds to MINOS.
Also shown in the figure are the regions at present excluded by E776  data
(shaded area) and the favoured range on $\Delta M^2_{DM}$ from dark matter
considerations (solid horizontal lines). For comparison also the  LSND data are
shown.}
\label{mutaudm}
\end{figure}

In Fig.\ \ref{mutaudm} we show the regions accessible in the
$(\sin^2(2\theta_{e\tau}),\Delta M^2_{DM})$ plane to the different experiments
for an optimum value of the $e\tau$ mixing $\sin^2(2\theta_{e\tau})=0.16$. As
seen in the figure, the whole parameter space accessible to the  CERN
experiments
CHORUS and NOMAD is already ruled out by the $E776$ data.  E803 and  NAUSICAA
at CERN can  marginally see this oscillation in the favoured range for $\Delta
M^2_{DM}$ for maximum $e\tau$ mixing. In particular if the LSND data are
confirmed, this oscillation could be  observable at both experiments for this
optimum value of the $e\tau$ mixing. For smaller values of the $e\tau$ mixing
these oscillations became very  marginal or invisible. NAUSICAA at Fermilab
can observe this oscillation  for
$\sin^2(2\theta_{e\mu})\sin^2(\theta_{e\tau})\gtrsim (0.3$--$1) \times
10^{-4}$. If the LSND results are confirmed, the same oscillation  could then
be detected  in NAUSICAA at Fermilab, provided  $\sin^2(\theta_{e\tau})\gtrsim
10^{-2}$. As seen in the figure the accessible  range for MINOS is already
excluded by the E776 data.

\begin{figure}
\begin{center}
\mbox{\epsfig{file=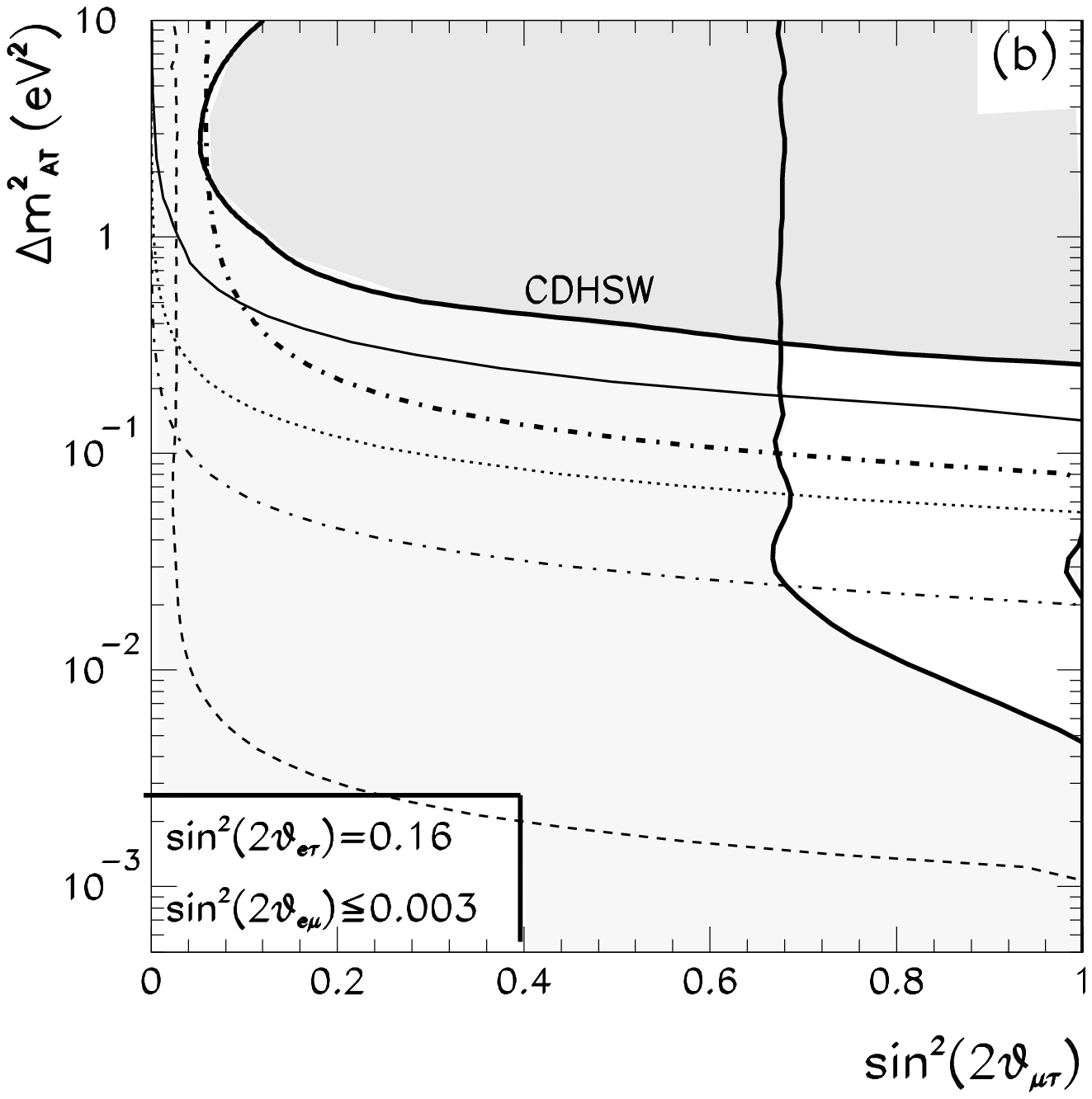,bbllx=60,bblly=300,bburx=600,bbury=700}}
\end{center}
\caption{{\bf (a) }
Accessible regions (90\% CL) for the $\nu_\mu\rightarrow \nu_\tau$  oscillation
in the ($\Delta m^2_{AT}$, $\sin^2(2\theta_{\mu\tau})$) plane  for zero value
of the mixing $e\tau$. The fine solid line corresponds  to the CHORUS/NOMAD
experiments and the dotted line corresponds to the Fermilab experiment E803 and
to the improved CERN experiment. The dot-dashed line corresponds to  NAUSICAA
at Fermilab and the dashed line to MINOS. Also shown in the figure are the
regions at present excluded  by CDHSW data  (dark shaded area) and the
atmospheric neutrino analysis (light shaded area).\\
{\bf (b)} Same as {\bf (a)} for maximum mixing $e\tau$.  Also shown in the
figure are the region at present excluded  by CDHSW  data (dark shaded area)
and the atmospheric neutrino analysis (light shaded area). These are the only
constraints for  zero or values of the $e\mu$ mixing $\sin^2(2\theta_{e\mu})\ll
0.003$. For maximum value  $\sin^2(2\theta_{e\mu})\simeq 0.003$ the E776 limit
(thick dot-dashed line) is relevant.}
\label{mutauat}
\end{figure}

Figure \ref{mutauat} shows the region accessible to the experiments in the
$(\sin^2(2\theta_{\mu\tau}), \Delta m^2_{AT})$ parameter  space for different
values of the other mixings. As seen in the figure, the possibility of
observing this oscillation  depends on the values of the
$e\tau$ and $e\mu$ mixings. For zero $e\tau$ mixing, the oscillation can be
observed by all experiments in the region allowed by all laboratory and
atmospheric neutrino experiments (see Fig.\ \ref{mutauat}.a).  It must be
pointed out, however, that  CERN experiments can observe this oscillation very
marginally only.   For larger values of the $e\tau$ mixing the situation is
almost
the same as long as the $e\mu$ mixing remains zero. If both mixings  take their
maximum allowed value the whole parameter space for the CERN experiments is
ruled out by the atmospheric and E776 experiments data  (see Fig.
\ref{mutauat}.b).

E803 and NAUSICAA at CERN can observe oscillations  with $\Delta m^2_{AT}
\gtrsim 5\times 10^{-2}$eV$^2$ for any allowed value  of the $e\tau$ and $e\mu$
mixings. NAUSICAA operated at the Fermilab beam could access oscillations with
$\Delta m^2_{AT}\gtrsim (2 \times 10^{-2})$ eV$^2$. \footnote{There is even the
possibility \cite{mishra} that NAUSICAA can be operated at two points, near the
beam and at a few tenths of km allowing a limit $\Delta m^2_{AT}\gtrsim (3
\times 10^{-3})$ eV$^2$ sufficient to close the atmospheric window}. MINOS  can
reach mass differences  as low as $\Delta m^2_{AT}\gtrsim (10^{-3})$  eV$^2$,
closing down the allowed window for the atmospheric neutrino experiments.

\section{Conclusion}
\label{conclu}
In this paper we have studied the discovery
potential of laboratory experiments searching for the oscillation
$\nu_\mu(\nu_e) \rightarrow \nu_\tau$, after considering existing data on solar
and atmospheric neutrino experiments as well as the results from laboratory
experiments on neutrino oscillations.  To understand all these results in a
common framework, we have introduced a  fourth sterile neutrino and we have
performed a comprehensive reanalysis of the data in this four-neutrino
scenario. As an outcome of this analysis we can predict the number of expected
events at future $\nu_\tau$ oscillations  experiments at  CERN and Fermilab for
the allowed oscillation parameters.  We find that at these facilities
$\nu_e\rightarrow \nu_\tau$ oscillations are dominated by mass differences of
the order of $\Delta M^2_{DM} \simeq 4$--$10$ eV$^2$  while $\nu_\mu\rightarrow
\nu_\tau$ oscillations could occur with both  $\Delta M^2_{DM}$  and  $\Delta
m^2_{AT}\simeq 0.3$--$10^{-3}$ eV$^2$.

Our predictions are rather pessimistic for the upcoming  experiments NOMAD and
CHORUS, which, we find, are able  to explore only a small area of the
oscillation parameter space in both $\nu_e\rightarrow \nu_\tau$ and
$\nu_\mu\rightarrow \nu_\tau$ channels. The prospects are much better for
future $\nu_\tau$ experiments. E803 and/or an improved detector (NAUSICAA)
at CERN can explore
$\nu_e\rightarrow \nu_\tau$ with $\sin^2(2\theta_{e\tau})\gtrsim 10^{-2}$.
As for $\nu_\mu\rightarrow \nu_\tau$,  the oscillations with  $\Delta M^2_{DM}$
could
be observed at E803/NAUSICAA-CERN in the range favoured by the LSND results,
while oscillations with $\Delta m^2_{AT}$ are accessible for mixing values
allowed by the atmospheric neutrino data if $\Delta m^2_{AT} \gtrsim 5\times
10^{-2}$ eV$^2$.

The proposed long-baseline experiment MINOS at Fermilab has the unique
potential of completely exploring the region dominated by $\Delta m^2_{AT}$
thus, if present atmospheric neutrino data are correct MINOS $must$ observe
the oscillation $\nu_\mu \rightarrow \nu_\tau$. Finally, a future
high-performance detector such as NAUSICAA at Fermilab could have the potential
of exploring a very sizeable region of the parameter space for all the
oscillation modes. The $\nu_e \rightarrow \nu_\tau$, oscillations could be
detected
in the limit $\sin^2(2\theta_{e\tau})\sim 10^{-3}$ and  $\Delta M^2_{DM} \sim
10$ eV$^2$; $\nu_\mu \rightarrow \nu_\tau$ oscillations dominated by $\Delta
M^2_{DM}$ could be detected in the limit $\sin^2(2\theta_{e\mu})\sim 3 \times
10^{-4}$ and  $\Delta M^2_{DM} \sim 10$ eV$^2,$ and $\nu_\mu \rightarrow
\nu_\tau$ oscillations dominated by  $\Delta m^2_{AT}$ could be detected in the
limit $\Delta m^2_{AT} \sim 2 \times 10^{-2}(\sim 3 \times 10^{-3})$
and  maximal mixing. Indeed,
NAUSICAA has the potential to observe $both$  $\nu_\mu \rightarrow \nu_\tau$
and $\nu_e \rightarrow \nu_\tau$ oscillations.

\acknowledgements
We want to thank P. Lipari for providing us with the
atmospheric neutrino fluxes. We acknowledge discussions with R.\ Vazquez, P.\
Hernandez,  R.\ Sundrum and S.\ R.\ Mishra. J.\ J. G.\ -C.\ acknowledges the
hospitality of Harvard University.

\begin{table}
\caption{Characteristic values of $L/E$
(m/MeV) for the different experiments considered in the analysis.}
\label{tab:lovere}
\begin{displaymath}
\begin{array}{|c|c|}
\hline
\mbox{Experiment} &   \langle L/E\rangle  \mbox{(m/MeV)} \\
\hline
\mbox{Solar neutrino}  & \sim 10^{10} \\
\mbox{Atmospheric neutrino} &  \sim 10^{2} \\
\mbox{Bugey}  &  \simeq 20 \\
\mbox{CDHSW} &  \simeq 0.8\\
\mbox{E776} &  \simeq 0.4 \\
\mbox{E531} &  \simeq 0.04 \\
\mbox{CHORUS/NOMAD} & \simeq 0.02 \\
\mbox{E803} & \simeq 0.01 \\
\mbox{MINOS} &  \simeq 70 \\
\hline
\end{array}
\end{displaymath}
\end{table}
\begin{table}
\caption{Summary of the performance parameters for CHORUS and NOMAD.}
\label{tab:nomad}
\begin{tabular}{|c|c|c|}
  & NOMAD & CHORUS  \\
\hline   Target & Drift chamber panels & Emulsion \\
Mass (t) & $\sim 3$ & 0.8 \\
Target thickness (in $X_0$) & $\sim 1$ & $\sim 4$ \\
Initial run period (yr) & $ 3$ & 3  \\
Number of CC Interactions & $ 1.1 \times 10^{6}$ & $  5 \times 10^{5}$  \\
$\tau$ identification & & \\
Technique & Pure kinematics & Pure vertex \\
Total efficiency in \% $(EFF\times BR)$ & $\sim 5$ & $\sim 10$  \\
Sensitivity: & & \\
$P_{e\tau}$ & $1.3 \times 10^{-2}$ & $0.8 \times 10^{-2}$   \\
$P_{\mu\tau}$ & $2\times 10^{-4}$ & $1.4\times 10^{-4}$  \\
\end{tabular}
\end{table}

\begin{table}
\caption{Summary of the performance parameters for E803 and MINOS.}
\label{tab:e803}
\begin{tabular}{|c|c|c|}
 & E803 & MINOS  \\ \hline
Mass (t) & $\sim 0.75$ & 10,000 \\
Target thickness (in $X_0$) & $\sim 3$ & Very Large\\
Initial run period (yr) & $ 4$ & 2  \\
Number of CC interactions & $\sim 9 \times 10^{6}$ & $ \sim 10^{5}$  \\
$\tau$ identification   &   & \\
Technique  & Pure vertex & Disappearance experiment (ratio NC/CC) \\
Total efficiency in \% $(EFF\times BR)$ & $\sim 10$ & $\sim 70$  \\
Sensitivity:  &                    &       \\
$P_{e\tau}$   &   $7 \times 10^{-4}$ & --   \\
$P_{\mu\tau}$ & $1.4 \times 10^{-5}$ & 0.012   \\
\end{tabular}
\end{table}

\begin{table}
\caption{Summary of the performance parameters for NAUSICAA}
\label{tab:nausicaa}
\begin{tabular}{|c|c|c|}
 & NAUSICAA (CERN) & NAUSICAA (FNAL) \\ \hline
Target & Silicon/Graphite & Silicon/Beryllium \\
Mass (t) & $\sim 1$ & $\sim 2$ \\
Target thickness (in $X_0$) & $\sim 1$ & $\sim 1$ \\
Initial run period (yr) & $ 4$ & 4  \\
Number of CC interactions & $ 1.3 \times 10^{6}$ & $ 2.4 \times 10^{7}$  \\
$\tau$ identification & & \\
Technique & Vertex and kinematics & Vertex and kinematics \\
Total efficiency in \% $(EFF\times BR)$ & $\sim 20$ & $\sim 10$  \\
Sensitivity: & & \\
$P_{e\tau}$ & $7 \times 10^{-4}$ & $ 7 \times 10^{-5}$   \\
 $P_{\mu\tau}$ & $2 \times 10^{-5}$ & $2 \times 10^{-6}$  \\
\end{tabular}
\end{table}

\protect
\protect
\begin{figure}
\begin{center}
\mbox{\epsfig{file=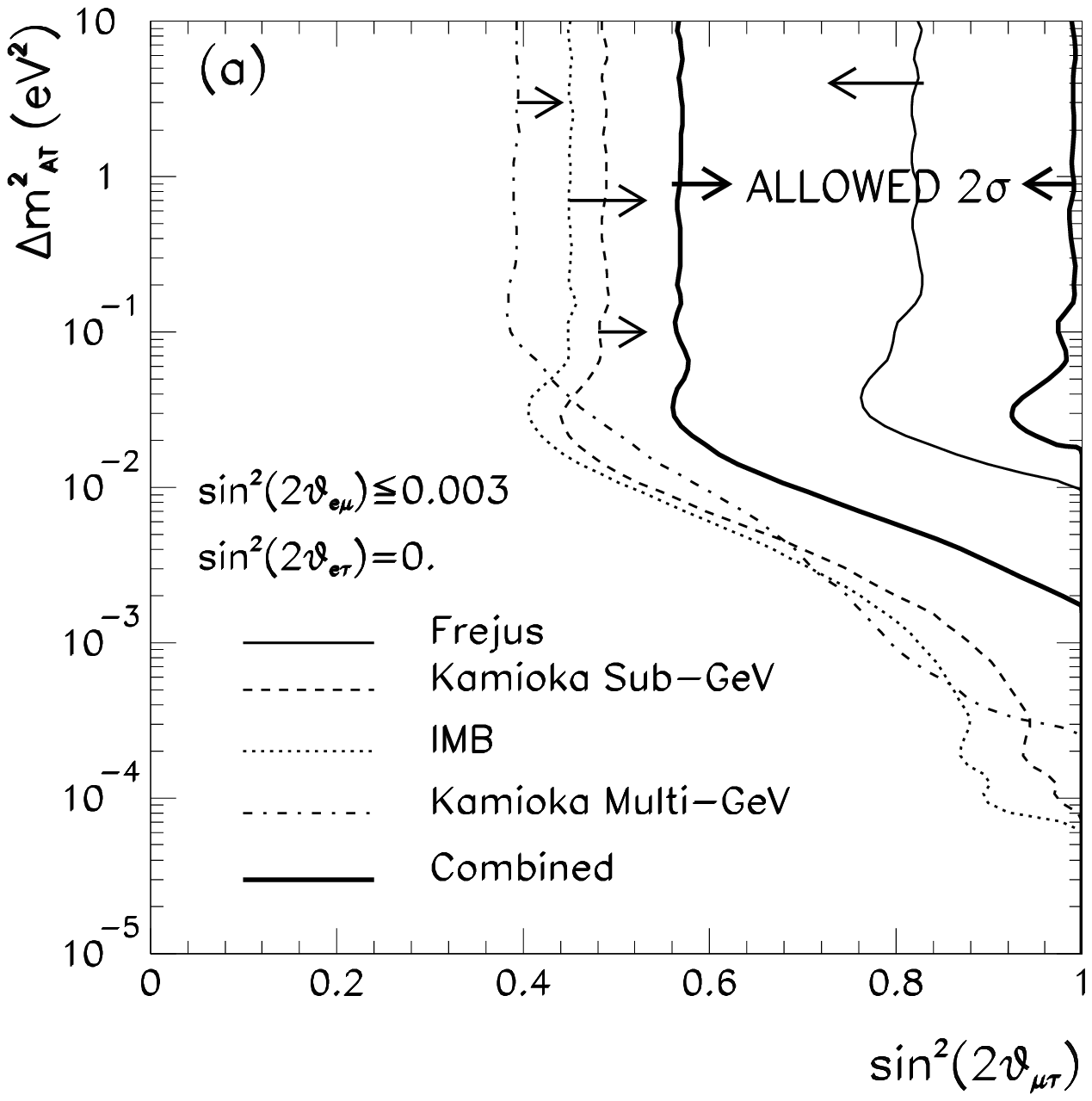,bbllx=60,bblly=200,bburx=600,bbury=600}}
\end{center}
\end{figure}
\protect
\begin{figure}
\begin{center}
\mbox{\epsfig{file=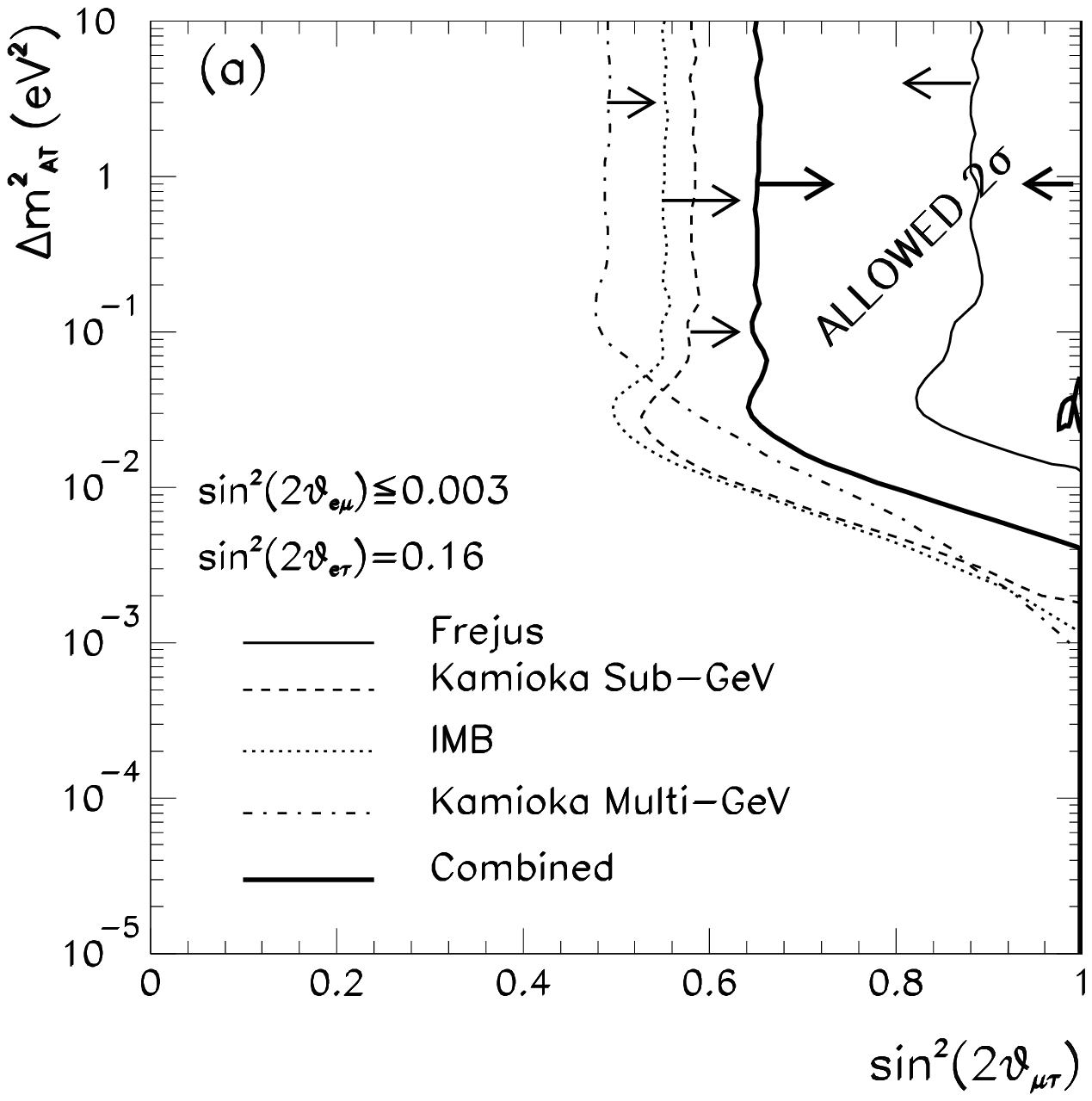,bbllx=60,bblly=200,bburx=600,bbury=600}}
\end{center}
\end{figure}
\begin{figure}
\begin{center}
\mbox{\epsfig{file=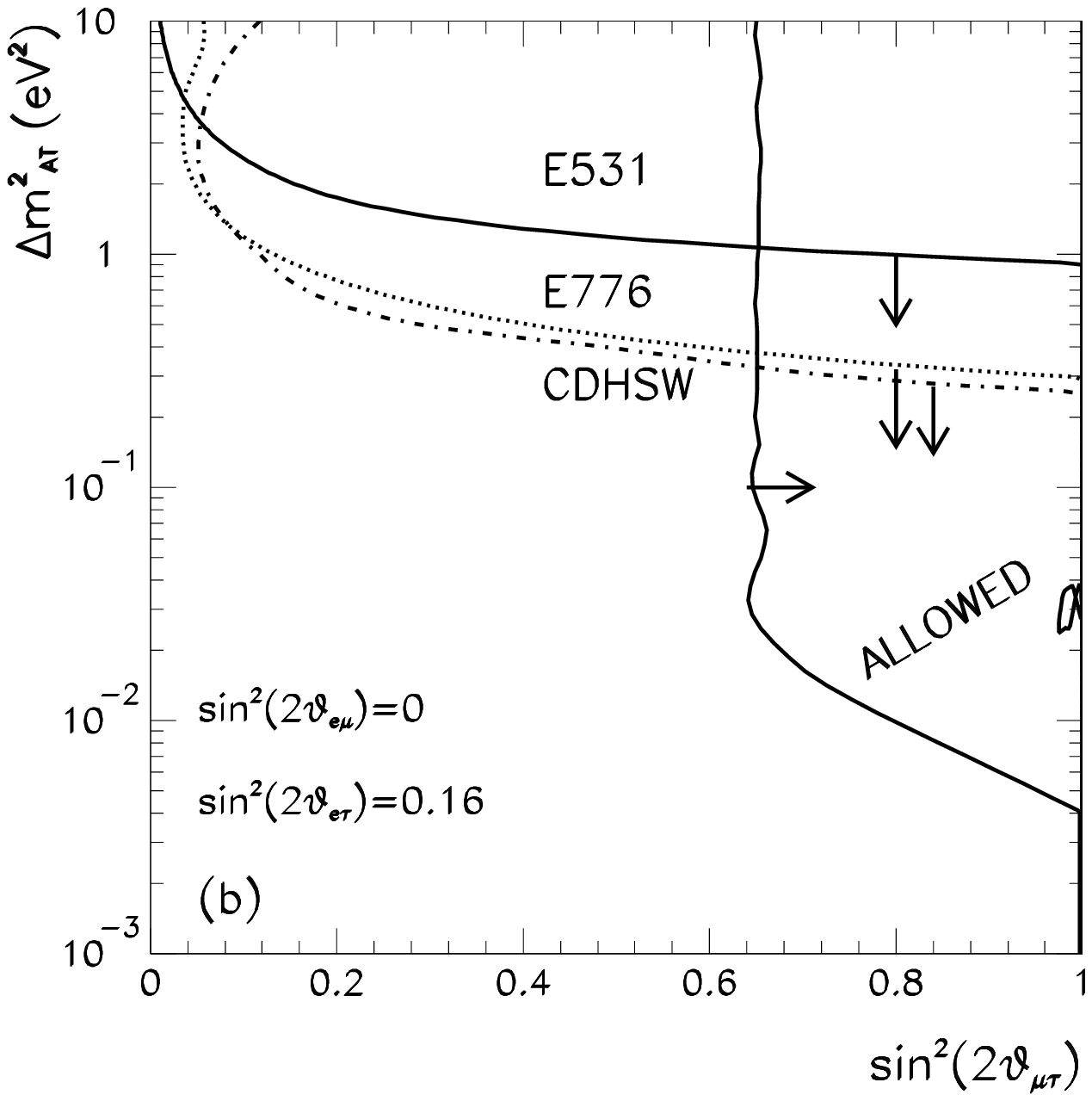,bbllx=60,bblly=200,bburx=600,bbury=600}}
\end{center}
\end{figure}
\protect
\protect
\protect
\protect
\begin{figure}
\begin{center}
\mbox{\epsfig{file=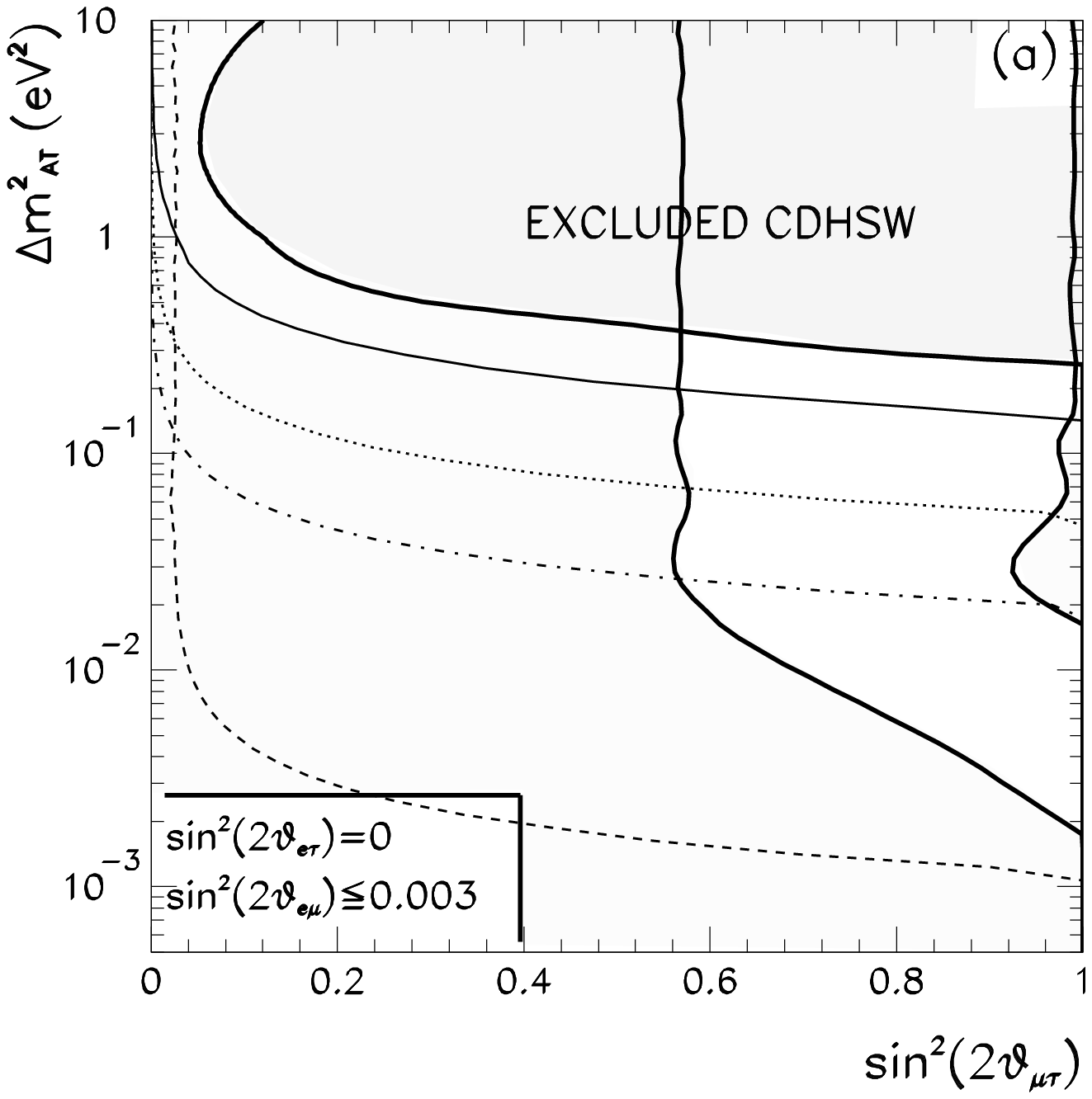,bbllx=60,bblly=300,bburx=600,bbury=700}}
\end{center}
\end{figure}
\end{document}